Submitted to **ADVANCED MATERIALS**

# A new diamond biosensor with integrated graphitic microchannels for detecting quantal exocytic events from chromaffin cells

By *F. Picollo\* ^, S. Gosso^, E. Vittone, A. Pasquarelli, E. Carbone, P. Olivero, V. Carabelli*

*^ These two authors contributed equally to this work*

[*]     Federico Picollo, Ettore Vittone, Paolo Olivero
        Department of Physics, NIS Centre of Excellence, CNISM Research Unit - University
        of Torino, INFN Sez. Torino
        via P. Giuria 1, Torino, 10125 (Italy)
        E-mail: picollo@to.infn.it

        Sara Gosso, Emilio Carbone, Valentina Carabelli
        Department of Drug Science and Technology, NIS Centre of Excellence, CNISM
        Research Unit - University of Torino
        Corso Raffaello 30, Torino, 10125 (Italy)

        Alberto Pasquarelli
        Institute of Electron Devices and Circuits, Ulm University
        Albert Einstein Allee 45, Ulm, 89069 (Germany)



*Abstract*

The quantal release of catecholamines from neuroendocrine cells is a key mechanism which

has been investigated with a broad range of materials and devices, among which carbon-based

materials such as carbon fibers, diamond-like carbon, carbon nanotubes and nanocrystalline

diamond. In the present work we demonstrate that a MeV-ion-microbeam lithographic

technique can be successfully employed for the fabrication of an all-carbon miniaturized

cellular bio-sensor based on graphitic micro-channels embedded in a single-crystal diamond

matrix. The device was functionally characterized for the *in vitro* recording of quantal

exocytic events from single chromaffin cells, with high sensitivity and signal-to-noise ratio,

opening promising perspectives for the realization of monolithic all-carbon cellular biosensors.





*Introduction*

The quantal release of bioactive molecules from neurons and neuroendocrine cells is a fundamental mechanism that regulates synaptic transmission and hormone release. In particular, chromaffin cells of the adrenal gland, expressing voltage-gated $Ca^{2+}$ channels functionally coupled to the secretory apparatus, represent an ideal system to study the exocytotic release of catecholamines (adrenaline, noradrenaline) from chromaffin granules, where they are stored at high concentration (1 M) together with ATP, opioids, peptides[1]. Carbon fiber microelectrodes (CFEs) are employed since few decades to detect the exocytotic activity of single excitable cells with amperometric techniques and provide significant information on key mechanisms such as the formation of the fusion pore and the kinetics of single secretory events in real time[2-5]. These classical carbon-based probes possess high, chemical stability and biocompatibility but can be hardly integrated in a miniaturized multi-electrode device. This limits the possibility of using them in multiple single-cell recordings or even to resolve secretory events within micro-areas of a single secretory cell. Two technical advances that would be beneficial for investigating the molecular basis of synaptic transmission in neuronal networks and the subcellular arrangement of the secretory apparatus.

The accessibility to a graphitic phase in single-crystal diamond through spatially-resolved lattice damage by means of energetic ion beams offers several appealing applications in microdevices fabrication. In particular, a technique based on MeV-ion beam lithography through variable-thickness masks was recently developed, which allowed the direct fabrication of buried graphitic micro-channels in single-crystal diamond at variable depth[6-9].

In this work we demonstrate that ion-beam lithography can be successfully employed for the fabrication of a monolithic all-carbon miniaturized cellular biosensor (μG-SCD) for recording the exocytotic activity of single chromaffin cells, with detection performances comparable to CFEs.





*Experimental*
*Device microfabrication*

The biosensor was realized with a synthetic single crystal diamond of $3\times3\times1.5$ mm$^3$. The diamond was produced with high pressure high temperature (HPHT) technique by Sumitomo Electrics and it is classified as type Ib, with a substitutional nitrogen concentration comprised between 10 and 100 ppm. The sample is cut along the (100) crystal direction and it is optically polished on the two opposite large faces. The crystal has good transparency in the visible spectrum, with an absorption band in the blue-violet region.

Sub-superficial conductive micro-paths were realized in the diamond matrix by means of a deep ion beam lithography technique based on direct focused ion beam writing through suitable variable-thickness masks, thus allowing for the modulation of the depth at which the channels are formed and therefore their emergence at specific locations of the sample surface. The fabrication method is described in details in previous works.[6-9] The above-mentioned variable-thickness masks were realized on the diamond surface by thermal evaporation of Cu. As shown schematically in Figure 1a, this arrangement allowed the deposition of two Cu 3-5 μm thick masking structures with slowly degrading edges. The diamond sample was then irradiated with a 1.8 MeV He$^+$ beam at the ion microbeam line of the AN2000 accelerator of the Legnaro National Laboratories.[10]

The energy transferred by ions to the diamond lattice induces the displacement of carbon ions and hence the creation of defects. Their distribution along the ion track follows a typical depth profile, with a peak at the end of range. In Figure 1a the damage density profile is reported: damage is parametrized in terms of vacancy density and derived from the SRIM-2008.04 Monte Carlo code[11] by taking an atom displacement energy value of 50 eV.[12] The vacancy density is obtained with a simple linear approximation by multiplying the output of SRIM simulations (number of vacancies per ion per unit depth) with the implantation fluence. Such





a crude approximation does not take into account complex non-linear damage effects such as defect-defect interaction and self-annealing[13] and therefore leads to a significant over-estimation of the vacancy density at high implantation fluences. Crossing the slowly-degrading metallic mask, the ions progressively lose energy and the highly damaged peak shallows. This allows the two terminals of the buried channel to emerge at the diamond surface, as shown schematically in Figure 1a.

After ion implantation, a thermal annealing was performed at a temperature of 1100 °C for two hours in high vacuum, allowing the recovering of the diamond structure of the low-damaged regions, and the conversion to graphite of those regions characterized by a vacancy density exceeding the graphitization threshold (which values comprised between $3 \cdot 10^{22}$ and $9 \cdot 10^{22}$ vacancies $cm^{-3}$). As a consequence, a 47-$\mu$m-wide, 1.6-mm-long and ~450-nm-thick graphitic channel was formed at a depth of 3 $\mu$m in the diamond matrix (see Fig 1b).

The graphitic channel was electrically characterized. The linear trend of the I-V curve (not reported) indicates an ohmic conduction occurring in the channel with a resistance of 2.7 k$\Omega$. The resistivity of the conductive structure was calculated using the above-mentioned channel dimensions. The obtained resistivity value ($\rho$~3.6 m$\Omega$ cm), is comparable with what reported for common polycrystalline graphite ($\rho$~1.3 m$\Omega$ cm[14]). . The sample was then assembled onto a carrier board (Roger R4003) provided with a 200 $\mu$l glass ring subsequently employed as perfusion chamber. one of the graphitic channel terminals was contacted with silver paste and connected to the electronic acquisition system together with an Ag/AgCl reference electrode immersed in the solution. Both the Ag electrode and the bonding wire were then passivated with sylgard, as shown schematically in Figure 1c. The complete assembly was mounted on a transmission microscope stage which was equipped with motor-driven micromanipulators. The above-mentioned micromanipulators were employed for both the cell fine positioning on the graphitic microelectrode with a glass patch-clamp pipette and for the reference amperometric measurement with Carbon Fiber Microelectrodes (CFEs).





*Isolation and culture of mouse adrenal medulla chromaffin cells*

All experiments were conducted in accordance with the guidelines on Animal Care established by the Italian Minister of Health and were approved by the local Animal Care Committee of Turin University.

Mouse chromaffin cells (MCCs) were obtained from young C57BL/6J male mice (Harlan, Milano, Italy).[15] After removal, the adrenal glands were placed in $Ca^{2+}$ and $Mg^{2+}$ free Locke's buffer, containing (in mM): 154 NaCl, 3.6 KCl, 5.6 $NaHCO_3$, 5.6 glucose and 10 HEPES (pH 7.3, at room temperature). The glands were then decapsulated in order to separate the medullas from the cortical tissue. Enzymatic digestion of the medulla was achieved by keeping the medulla for 20 min at 37 °C into a DMEM solution enriched with 0.16 mM Lcysteine, 1 mM $CaCl_2$, 0.5 mM EDTA, 20 U $ml^{-1}$ of papain (Worthington Biochemical, Lakewood, NJ, USA), 0.1 mg $ml^{-1}$ of DNAse (Sigma, Milan, Italy). The digested glands were then washed with a solution containing DMEM, 1 mM $CaCl_2$, 10 mg $ml^{-1}$ BSA and resuspended in 2 ml DMEM supplemented with 15% fetal bovine serum (FBS) (Invitrogen, Grand Island, NY, USA). Isolated chromaffin cells were obtained after mechanical disaggregation of the glands. Cells were then incubated at 37°C in a water-saturated atmosphere with 5% $CO_2$ and used within 2-4 days after plating. Cells were maintained in culture using non-adherent dishes, pretreated using 5% BSA.

*Cyclic voltammetry measurements*

With the aim of investigating the dark current of the electrode, the redox properties of the saline solution and the voltage corresponding to the oxidation peak of adrenaline, cyclic voltametric measurements were performed to characterize the microelectrode material in the working environments. To this purpose, two solutions were used: the physiological external





solution containing (in mM): 128 NaCl, 2 MgCl$_2$, 10 glucose, 10 HEPES, 10 CaCl$_2$ and 4 KCl, and a solution of adrenaline (0.1 M) diluted in distilled water.

A voltage from -1 V to +1.5 V was applied to the graphitic micro-electrode with respect to the reference electrode, with a scan rate of 50 mV s$^{-1}$. The current collected at the micro-electrode for different bias voltages was recorded and processed using a home-developed electronic acquisition front-end and a 16-bit National Instrument digitizer with a sampling rate of 4 kHz and bandwidth from DC to 1 kHz.[16]

As shown in Figure 2, when the μG-SCD device was immersed in the above-described physiological saline solution, no redox activity could be detected within the anodic range of the hydrolysis window, with a leakage current lower than 10 pA up to a bias voltage of 1 V. In presence of the above-mentioned adrenaline solution, the graphitic electrode exhibited a catalytic activity corresponding to the oxidation of the molecule at the biased micro-electrode, as shown in the red-line plot of Figure 2. This curve allows the evaluation of the optimal bias voltage for the subsequent amperometric measurement, which was set to +800 mV, corresponding to the maximum value of the (oxidation current):(water hydrolysis) ratio.

*Exocytosis detection from chromaffin cells*

The quantal release of adrenaline from mouse adrenal medulla chromaffin cells was monitored by keeping the graphitic micro-electrode at +800 mV bias with respect to the reference electrode. The same acquisition system was used for amperometric measurements by means of standard 5 μm diameter carbon fiber microelectrodes (CFEs), positioned in close proximity to the cell membrane and biased at +800 mV. Chromaffin cells were suspended in the above-described physiological solution then loaded into the device reservoir. Cells were placed onto the microelectrode by means of a patch-clamp pipette controlled by a motor-driven micromanipulator under visual inspection through the inverted microscope. The cell was lowered until it touched the diamond substrate in correspondence of the graphitic





electrode. As previously shown, $Ca^{2+}$ channels and secretory responses are not impaired by floating culture conditions.[16] In 2-min recordings the current signal was sampled with an EPC-10 amplifier (HEKA) and a digitizing system with sampling rate 4 kHz and low-pass-filtered at 1 kHz . The background amperometric current had 5 pA peak-to-peak noise, as shown in Figure 3a.

Exocytosis was stimulated by applying 50 µl of a KCl-enriched solution (in mM): 100 NaCl, 2 $MgCl_2$, 10 glucose, 10 HEPES, 10 $CaCl_2$, 30 KCl. Sequences of amperometric spikes started after cell stimulation by means of the KCl solution. Typical recordings of the stimulated exocytosis using the graphitic microchannels and standard CFEs are shown in Figure 3b and 3c, respectively. In the insets typical time evolutions of well-resolved single amperometric spikes are reported. The amplitude of most amperometric spikes was well above the background noise level (5 pA peak-to-peak) and varied from 8 to 180 pA.

Systematic data analysis of amperometric spikes was performed by means of "Igor" macros, as previously described.[17] Spikes with amplitudes below 8 pA were discarded from the analysis. As indicated in Table I, amperometric spikes detected by the graphitic microelectrodes (n = 95) were comparable with signals obtained by conventional CFEs (n = 505).[16] As reported in Table I, the mean current amplitudes measured by the graphitic microelectrodes and CFEs were respectively (26 ± 4) pA and (33 ± 6) pA (p > 0.05), while the integrated charges were respectively (0.28 ± 0.04) pC and (0.35 ± 0.07) pC (p > 0.05). In Table I also the values of $m$ (i.e. the slope of the rising phase of the amperometric spike, according to a linear approximation) and of $t_{1/2}$ (i.e. f the half-height fall time of the amperometric spike) are reported. Data obtained with the diamond-embedded graphitic microelectrode are compatible within the statistical distributions with the signals obtained with conventional CFEs,[16] thus demonstrating the suitability of a diamond-based miniaturized device for the recording with high sensitivity and signal-to-noise ratio of





extracellular quantal signals in an all-carbon multi-electrode arrangement which cannot be implemented with conventional CFEs.

### *Discussion and conclusions*

The monolithic fabrication with MeV ion-beam lithography of buried graphitic microchannels in single-crystal diamond for applications in single-cell sensing of exocytotic activity has been demonstrated, and the first functional tests of a μG-SCD device have been reported.

Buried conductive micro-channels were fabricated by means of direct writing with a scanning MeV ion-beam, taking advantage of the possibility offered by a high-energy ion probe to locally amorphize and subsequently graphitize the diamond structure. In particular, variable thickness masks were adopted to modulate the penetration depth of the ions and hence allow the emergence of the conductive paths at the sample surface in correspondence of their endpoints.

The active substrate was mounted on a prototypical device that allowed the *in vitro* detection of quantal neurotransmitter release from single excitable cells kept in a physiological solution. The recorded cellular signals demonstrate that an all-carbon miniaturized device based on graphitic electrodes embedded in a diamond matrix can efficiently detect the quantal release of catecholamines from secretory cells. In particular, it is worth noting that the performance (amperometric sensitivity, signal-to-noise ratio, time resolution) of the prototypical μG-SCD device presented in this work compare well not only with standard CFEs (see Table I), but also with other well developed technologies at the state of the art, such as devices based on indium tin oxide (ITO) conductive glass,[18-20] noble metals (Au, Pt, …)[21-22] and boron-doped nanocrystalline diamond (B:NCD).[16] In conclusion, these results open promising perspectives for the realization of all-carbon multielectrode miniaturized devices in artificial diamond (a material which is becoming available with increasing crystal quality at ever-decreasing costs[23-24]) in which full advantage of the robustness, chemical stability,





biocompatibility and transparency can be exploited to obtain multiparametric signals detection from cell networks. This approach has the potential to drastically accelerate data collection from large ensembles of cells, leaves the experimental sample under physiological conditions. In addition, it is non-invasive and allows repeated measurements over time. Furthermore, geometry and transparency of the structures allows multi-techniques measurements, since other probes (i.e. patch-clamp pipettes) can be easily adapted.

((Supporting Information is available online from Wiley InterScience or from the author)).

Received: ((will be filled in by the editorial staff))
Revised: ((will be filled in by the editorial staff))
Published online: ((will be filled in by the editorial staff))

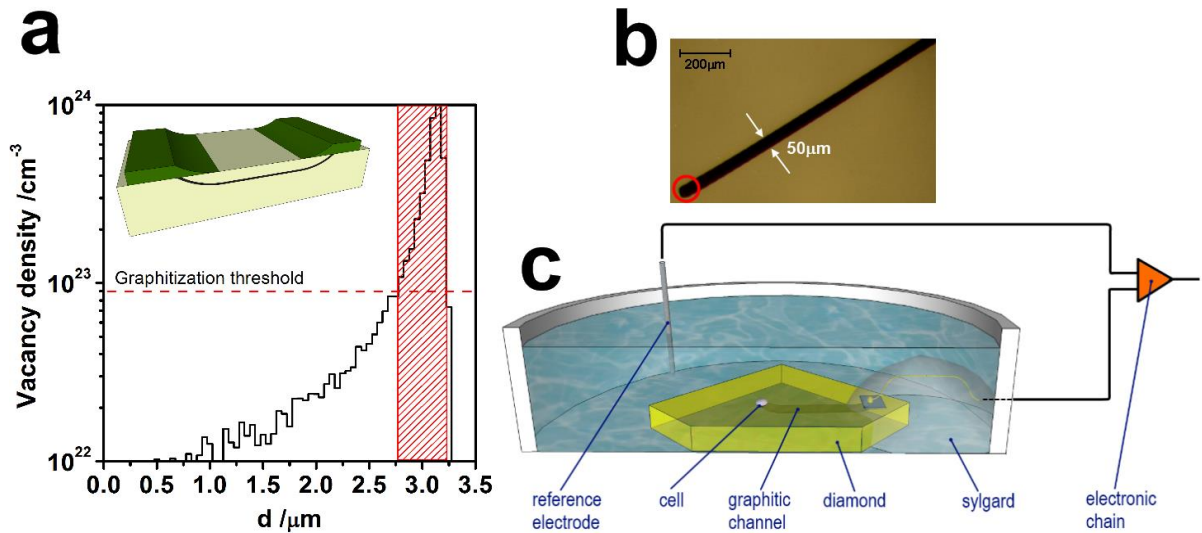

**Figure 1.** ( (a) Vacancy density profile induced by 1.8 MeV He[+] ions implanted in diamond at a fluence of $5 \cdot 10^{17}$ cm[-2]. The horizontal line indicates the graphitization threshold, while the patterned rectangle highlights the thickness of the graphitic layer formed upon thermal annealing, into the inset is represented a schematics of the fabrication of buried channels emerging at the diamond surface in proximity of the edges of the variable-thickness metallic masks. (b) Optical micrograph in transmission geometry of the diamond sample with the buried graphitic channel (black line). (c) Schematic representation of the device ready for the amperometric recording connected to the electronic acquisition system (not in scale).)





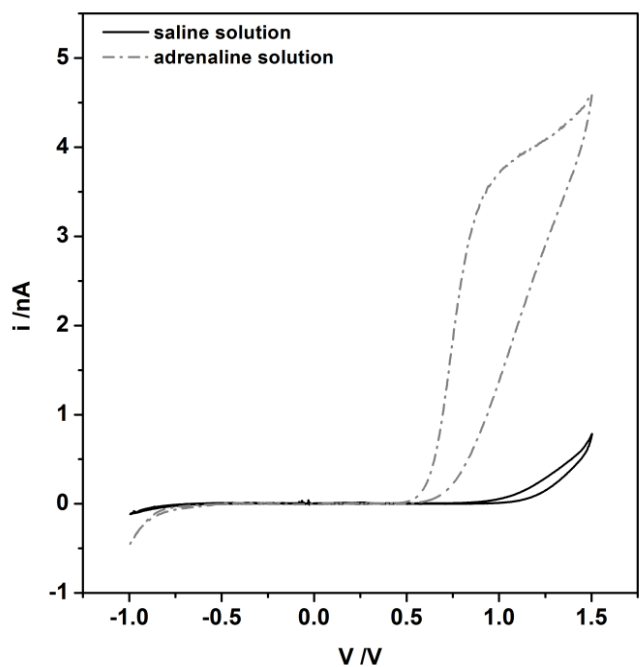

**Figure 2.** (Cyclic voltammetry measurements collected by keeping the device immersed in saline (blue line) and 0.1 M adrenaline (red line) solutions; the scan rate is 50 mV s⁻¹.)

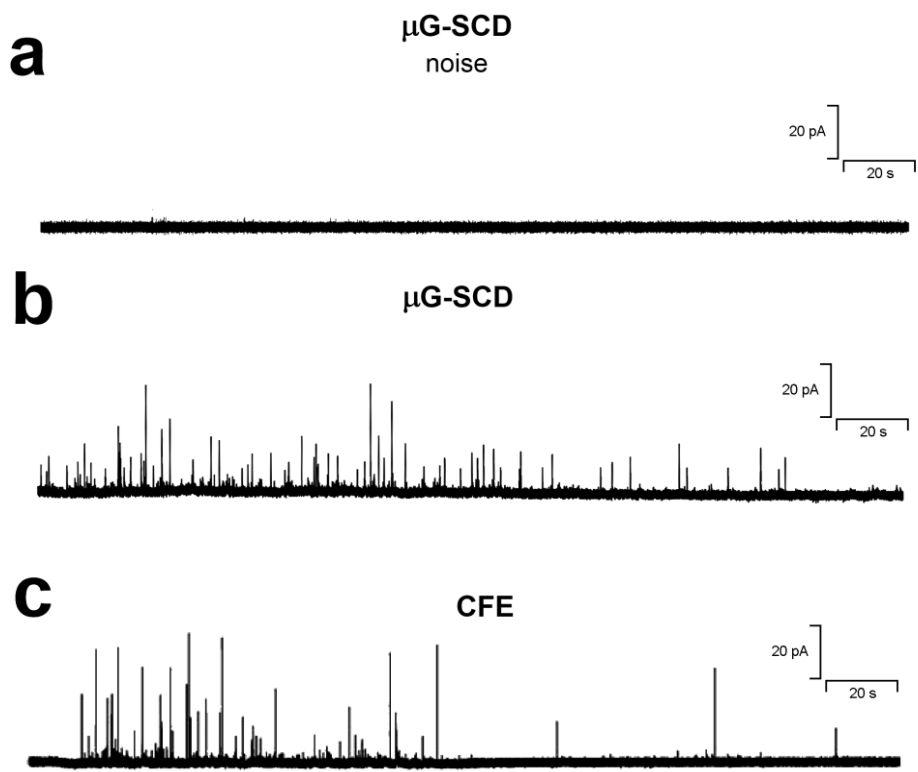

**Figure 3.** ((a) Background current measured from the graphitic microelectrodes, prior to stimulation of the chromaffin cell. Typical amperometric recordings from a chromaffin cell positioned in close proximity to the graphitic electrode at a +800 mV bias with respect to the reference electrode (b) and with standard CFEs (c), after stimulation with a KCl-enriched solution.)





**Table 1.** (Mean values and standard deviations of the main parameters describing the shape of amperometric spikes using graphitic microelectrode (94 spikes) compared with values obtained by means of standard CFEs.[15-16] Exemplificative amperometric spikes are report for both the techniques)

| | CFE | μG-SCD |
|---|---|---|
| | 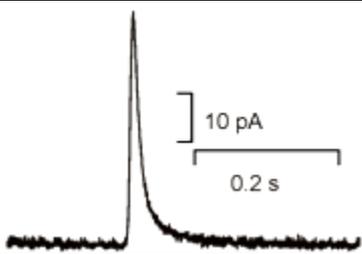 | 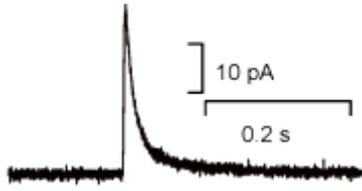 |
| $I_{max}$ [pA] | 33 ± 6 | 26 ± 4 |
| Q [pC] | 0.28 ± 0.04 | 0.35 ± 0.07 |
| $Q^{1/3}$ [pC$^{1/3}$] | 0.57 ± 0.02 | 0.61 ± 0.05 |
| m [nA s$^{-1}$] | 14 ± 2 | 15 ± 3 |
| $t_{1/2}$ [ms] | 9.4 ± 0.8 | 7.1 ± 0.8 |





This work was supported by: "FIRB - Futuro in Ricerca 2010" project (CUP code: D11J11000450001), funded by the Italian Ministry for Teaching, University and Research (MIUR); FENS-POR project "MicroDiBi", funded by the "BioPMed" scientific pole of Regione Piemonte; "Linea 1A ORTO11RRT5" projects of the University of Torino, funded by "Compagnia di San Paolo".